# Ultrafast Dynamics of Intrinsic Anomalous Hall Effect in the Topological Antiferromagnet Mn$_3$Sn

Takuya Matsuda[1]*, Tomoya Higo[2], Takashi Koretsune[3], Natsuki Kanda[1], Yoshua Hirai[2], Hanyi Peng[2], Takumi Matsuo[2], Naotaka Yoshikawa[2], Ryo Shimano[2,4], Satoru Nakatsuji[1,2,5,6], and Ryusuke Matsunaga[1,5]*

[1]*The Institute for Solid State Physics, The University of Tokyo, Kashiwa, Chiba 277-8581, Japan.*

[2]*Department of Physics, The University of Tokyo, Bunkyo-ku, Tokyo, 113-0033, Japan*

[3]*Deparment of Physics, Tohoku University, Sendai, 980-8578, Japan*

[4]*Cryogenic Research Center, The University of Tokyo, Bunkyo-ku, Tokyo, 113-0032, Japan*

[5]*Trans-scale Quantum Science Institute, The University of Tokyo, Bunkyo-ku, Tokyo, 113-0033, Japan*

[6]*Institute for Quantum Matter and Department of Physics and Astronomy, Johns Hopkins University, Baltimore, 21218 Maryland, USA*

*e-mail: matsuda.t@issp.u-tokyo.ac.jp, matsunaga@issp.u-tokyo.ac.jp

**Abstract**

**We investigate ultrafast dynamics of the anomalous Hall effect (AHE) in the topological antiferromagnet Mn$_3$Sn with sub-100 fs time resolution. Optical pulse excitations largely elevate the electron temperature up to 700 K, and terahertz probe pulses clearly resolves ultrafast suppression of the AHE before demagnetization. The result is well reproduced by microscopic calculation of the intrinsic Berry-curvature mechanism while the extrinsic contribution is clearly excluded. Our work opens a new avenue for the study of nonequilibrium AHE to identify the microscopic origin by drastic control of the electron temperature by light.**

**Main Text**

The anomalous Hall effect (AHE), a deflection of current flow perpendicularly to the bias electric field as a consequence of the broken time-reversal ($\mathcal{T}$) symmetry in magnets, has been one of the central issues in modern condensed matter physics because of its topological nature [1]. The intrinsic anomalous Hall conductivity $\sigma_{xy}^{\text{int}}$ in the DC limit is independent of scattering and solely determined by the band structure [2], as represented by

$$\sigma_{xy}^{\text{int}} = \frac{e^2}{\hbar} \sum_n \int \frac{\mathrm{d}\boldsymbol{k}}{(2\pi)^3} f(\epsilon_n(\boldsymbol{k})) B_{n,z}(\boldsymbol{k}), \qquad (1)$$

where $n$ is a band index, $f(\epsilon_n(\boldsymbol{k}))$ is an electron distribution function, and $B_{n,z}(\boldsymbol{k})$ is the Berry curvature in the momentum space. By contrast, the extrinsic AHE originating from the skew scattering [3] or side-jump mechanism [4] is induced by impurity scattering. To study the AHE, several approaches have been implemented to modulate a parameter for the anomalous transport. Because of the distinct dependences on scattering for the intrinsic and extrinsic mechanisms, the universal scaling law between the Hall and longitudinal conductivities can be used to discern the microscopic origins [5]. However, certain materials can deviate from the scaling [6]. Tuning the Fermi level by the gate voltage is also highly informative to investigate the AHE [7-9]. However, this method requires the gate electrode and is restricted for two-dimensional systems. In addition, the tuning range is limited especially for conducting materials. The element substitution can largely control the Fermi level [10], but concomitantly alters the magnetic properties and the scattering rate. As a non-contact method without using electrodes nor dopants, optical spectroscopy using Faraday or Kerr rotation has been utilized to discuss the origin of AHE [11] because the experimental result can be compared to the intrinsic Hall conductivity spectrum calculated from the band structure. To obtain the spectrum close to the DC transport, terahertz time-domain spectroscopy (THz-TDS) with the Faraday rotation is important because the response functions are obtained as complex quantities without the Kramers-Kronig transformation [12-21]. However, the comparison between the experiment and the first-principle calculation is difficult at the low energy scale of millielectronvolts.

As an alternative approach, we employ ultrafast pump-probe spectroscopy. Photoexcited electrons are rapidly scattered to form quasi-thermalized distribution with an elevated electron temperature of several hundred kelvins, while the lattice degree of freedom remains intact until the excess energy is transferred via the electron-phonon interaction. By using the electron temperature as a largely controllable parameter, the transient AHE away from equilibrium can be probed by another short THz pulse. Although the static THz Faraday rotation spectroscopy has been performed in many magnets [12-21], to the best of our knowledge, ultrafast time-resolved spectroscopy has been reported only in a ferrimagnet [14], where sub-100 ps dynamics was discussed. The similar time resolution was also recently achieved in the electrical detection [22] and very recently used to study switching dynamics of Mn$_3$Sn [23]. However, ultrafast dynamics in the femtosecond regime has evaded experimental investigation thus far.

As a target material, we choose a topological antiferromagnet Mn$_3$Sn. Despite the absence of net magnetization, Mn$_3$Sn shows large AHE near room temperature comparable with that in ferromagnets [24]. As shown in Fig. 1(a), the antiferromagnetic phase is characterized by the cluster-octupole moment on the *a-b* plane which explicitly breaks the $\mathcal{T}$ symmetry [25], resulting in the topologically nontrivial band structure [26]. The large AHE has opened a new avenue for novel antiferromagnetic spintronics [27,28]. Ultrafast dynamics of AHE in Mn$_3$Sn is to be clarified for realization of high-speed readout spintronic information. Furthermore, Mn$_3$Sn thin films show a peculiar spin-reorientation phase transition as shown in Fig. 1(b). While the cluster-octupole phase develops below $T_\mathrm{N}$~430 K, the Hall resistivity $\rho_{xy}$ starts to decrease below 220 K due to the transition into the helical phase [29] where the cluster-octupole moment rotates along the *c*-axis with mitigating a degree of the broken $\mathcal{T}$ symmetry. Consequently, the AHE near 150 K increases with increasing temperature, a trend that is opposite to that in usual magnets. Therefore, photoinduced change of the AHE at the cluster-octupole and helical phases is highly intriguing to discuss what happens after photoinduced increase in the electron temperature.

In this Letter, we report the first study of ultrafast time-resolved AHE in a Mn$_3$Sn thin film by using optical pump-THz Faraday rotation probe spectroscopy with a sub-100 fs time resolution, a few orders of magnitude faster than the earlier works. We observe a peculiar time trace of AHE, which clearly resolves that AHE is drastically suppressed by photoexcited electrons before demagnetization. The result agrees well with a microscopic view with the intrinsic mechanism, providing unprecedented insights into the nonequilibrium nature of AHE.

The sample is a polycrystalline Mn$_3$Sn thin film deposited on SiO$_2$ substrates by DC magnetron sputtering [30]. The film thickness of 20 nm is less than the penetration depth (22 nm) of the optical pump so that inhomogeneous excitation is avoided. The characterization of the sample is described in the Supplementary Material [31]. By using polarization-resolved THz-TDS [37], we evaluated the longitudinal and anomalous Hall conductivity spectra $\tilde{\sigma}_{xx}(\omega)$ and $\tilde{\sigma}_{xy}(\omega)$ at 280 K, as respectively shown in Figs. 1(d) and 1(e). $\tilde{\sigma}_{xx}(\omega)$ exhibits a typical Drude response with a scattering rate of ~10 fs [15,38]. $\text{Re}\tilde{\sigma}_{xy}(\omega)$ is flat and as large as 20 $\Omega^{-1}\text{cm}^{-1}$ and $\text{Im}\tilde{\sigma}_{xy}(\omega)$ is negligibly small, showing that the THz AHE is also within the DC limit [15]. The circles on the left axes are the data obtained in the DC transport with values lower than the THz conductivity. This can be attributed to the effects of grain boundaries or surface roughness in the DC transport in such a very thin sample [39].

Figure 1(c) shows a schematic of the pump-probe setup. The THz pulse was linearly polarized along the $x$-direction before entering the sample and detected by the THz-TDS with the transmission geometry. By changing the configuration of a wire-grid polarizer (WGP2), the $x$- and $y$-components of the THz field after transmittance by the sample can be obtained separately. By scanning the THz pulse delay time $t_{\text{probe}}$, the THz pulse waveform was evaluated and converted to the frequency domain. The optical pump pulses with 1.55-eV photon energy and 40-fs pulse duration irradiate the sample with a controllable pump delay $t_{\text{pump}}$. We performed the experiments under the magnetic field of +2 and -2 T and obtained $E_x$ and $E_y$ of the THz field as even and odd components for the magnetic field, respectively, and evaluated the longitudinal and Hall conductivity spectra $\tilde{\sigma}_{xx}(\omega)$ and $\tilde{\sigma}_{xy}(\omega)$ [31].

First, we conducted the pump-probe experiment at 220 K at which the AHE is maximized. Figure 2(a) shows $E_x$ and $E_y$ of the THz pulse waveform after transmittance by the sample without the pump. To detect the ultrafast dynamics, we set $t_{\text{probe}}$ equal to the peak ($t_{\text{probe}} = t_0$ in Fig. 2(a)) and scan $t_{\text{pump}}$ to detect the change in $E_y$ induced by the pump, $\delta E_y$. Figure 2(b) shows the results of $\delta E_y$ as a function of $t_{\text{pump}}$ with pump fluences $I_{\text{p}}$=300 and 500 µJ cm$^{-2}$. Immediately after the pump, $E_y$ rapidly decreased, and subsequently relaxed within a few picoseconds. We fitted the data and evaluated the rising times of $\delta E_y$ as 50 and 120 fs for $I_{\text{p}}$=300 and 500 µJ cm$^{-2}$, respectively, which are comparable to the time resolution limited by the pulse duration of 40 fs. To obtain the spectral information, we set $t_{\text{pump}}$ to 0.52 ps and scanned $t_{\text{probe}}$. Figure 2(c) shows $\text{Re}\tilde{\sigma}_{xy}(\omega)$ for different pump fluences. Averaging the data between 2.0 and 5.2 meV, we found that $\text{Re}\tilde{\sigma}_{xy}(\omega)$ decreases by 40% for $I_{\text{p}}$=500 µJ cm$^{-2}$. Moreover, we also investigated the change in $E_x$ and compared $\text{Re}\tilde{\sigma}_{xx}(\omega)$ with and without the pump in Fig. 2(d). We found that the change in $\tilde{\sigma}_{xx}(\omega)$ is negligibly small. A fitting with the Drude model suggested that the change in the scattering time by the pump is only 3% at most [31].

Previously photoexcited dynamics of magnets has been studied by using time-resolved magneto-optical Kerr effect (TR-MOKE) or magnetic circular dichroism, etc [40]. The polarization rotation of light $\theta(\omega)$ by the Faraday or Kerr effect is proportional to $\sigma_{xy}(\omega)$ as far as $\theta(\omega)$ is small. According to the Onsager's theorem, $\sigma_{xy}(\omega)$ is expressed as $\sigma_{xy}(\omega) = \alpha(\omega) \times M$, where $M$ is the magnetization and $\alpha(\omega)$ is a coefficient determined by the electron and lattice systems. Since the discovery of ultrafast TR-MOKE signal within 100-300 fs in ferromagnets [41,42], many efforts have been devoted to clarifying ultrafast demagnetization with possible spin-flip scattering mechanisms [43,44] and superdiffusive transport [45]. In this work we exclude the superdiffusive transport because the sample is thinner than the pump penetration depth and the probe size is as large as 6 mm so that the spin diffusion away from the probe spot is unlikely. In addition, it is to be noted that an ultrafast change of $\sigma_{xy}(\omega)$ does not necessarily indicate a change in the magnetization because of the relation:

$$\Delta\sigma_{xy} = \Delta\alpha \times M + \alpha \times \Delta M. \qquad (2)$$

Equation (2) means that even if the magnetization is unchanged ($\Delta M = 0$), the $\Delta\sigma_{xy}$ signal can be observed when the electron system largely changes ($\Delta\alpha \neq 0$) in magnets ($M \neq 0$). Because the electron system largely changes right after the photoexcitation, the first term $\Delta\alpha \times M$ can be dominant in the ultrafast regime [46-49]. However, the existence of the first term has been often neglected in many literatures. For thorough understanding of ultrafast dynamics, the electron contribution must be distinguished from the change of spins. Note that $M$ is a parameter that breaks the $\mathcal{T}$ symmetry and corresponds to the cluster-octupole moment in the case of Mn$_3$Sn.

To elucidate the nonequilibrium dynamics, we cooled the sample down to 150 K. Figure 3(a) shows the temperature dependence of the THz Hall conductivity, showing that the AHE near 150 K increases with increasing temperature in the helical phase in Fig. 3(b). Therefore, if the photoexcitation at 150 K heats the spin system, photoinduced "enhancement" of the cluster-octupole phase is expected to occur with an increase in $\sigma_{xy}$. The result of the pump-probe experiment at 150 K with $\delta E_y$ as a function of $t_{\text{pump}}$ is shown in Fig. 3(c). Interestingly, we observed that $\delta E_y$ first became *negative* with the timescale of ~100 fs similarly to the result at 220 K. A few picoseconds later, the signal in turn became *positive* by flipping its sign. By fixing $t_{\text{pump}}$ to 0.52 and 12 ps and by scanning $t_{\text{probe}}$, we obtained Re$\Delta\tilde{\sigma}_{xy}(\omega)$ in Figs. 3(d) and 3(e), respectively. The results clearly showed that the AHE is first *suppressed* and subsequently *enhanced*. By averaging the data between 2.0 and 5.2 meV at $t_{\text{pump}}$=12 ps, the AHE was increased by ~6 $\Omega^{-1}$cm$^{-1}$ for $I_{\text{p}}$=500 μJ cm$^{-2}$, corresponding to an increase in the temperature by ~40 K in thermal equilibrium.

We used the two-temperature model to evaluate photoinduced heating. Figure 4(a) shows the simulation results at the original temperatures of 220 and 150 K and $I_p$=500 µJ cm$^{-2}$ [31]. The optical pump excites electrons into higher energy bands, and these electrons are expected to immediately form a quasi-thermalized distribution with an electron temperature $T_e$. The maximum $T_e$ is as high as 700 and 630 K for the pump at 220 and 150 K, respectively. Subsequently, electron-phonon coupling elevates the lattice temperature $T_L$ such that the two systems are equilibrated within several picoseconds with an increase in the temperature of ~40 K, which is in good agreement with the increase in AHE observed in Fig. 3(e). Therefore, the positive sign of $\delta E_y$ around ~10 ps in Fig. 3(c) is attributed to pump-induced enhancement of the cluster-octupole phase as a result of heating the sample from 150 K. In contrast, the ultrafast suppression of Re$\tilde{\sigma}_{xy}(\omega)$ within 100 fs in Fig. 3(c) cannot be explained by the change in the spin configuration. The same argument also holds at 220 K, at which the pump would heat the sample at the timescale of ~10 ps and induce demagnetization. However, the temperature increase by 40 K has a minor effect on the AHE at 220 K as shown in Fig. 3(a) and therefore the demagnetization at ~10 ps is not clearly identified in Fig. 2(b). Importantly, the ultrafast reduction of Re$\tilde{\sigma}_{xy}(\omega)$ within 100 fs is not explained by heating the sample and therefore requires a more microscopic analysis.

Because the spin is a degree of freedom of electrons, an ultrafast change in the electron distribution itself may simultaneously alter the spin moment. To evaluate it, we calculated the band structure of Mn$_3$Sn by using density functional theory (DFT), as shown in Fig. 4(c). By considering finite electron temperatures, we calculated the local spin moment which is proportional to the cluster-octupole moment [25], as shown in Fig. 4(b). Note that a previous angular-resolved photoemission spectroscopy in a comparison with the DFT calculation revealed that the band structure of Mn$_3$Sn in the vicinity of $E_F$ is renormalized by a factor of 5 owing to strong many-body correlation [26]. Therefore, $T_e$=700 K corresponds to $T_{DFT}$=3,500 K in the calculation. Figure 4(b) shows that the increase from $T_e$=220 to 700 K ($T_{DFT}$=1,100 to 3,500 K) is accompanied by a decrease in the local spin moment, although only by 6%. Such insensitivity of the local spin moment to $T_e$ can be explained by the projected density of states for the cluster-octupole moment as shown in the right panel of Fig. 4(c) [25], showing that the spin configuration is determined mostly by the electrons of which the energy is a few eV below $E_F$. Therefore, the change in the electron distribution in the scale of $T_{DFT}$=3,500 K (~0.3 eV) does not immediately alter the spin moment.

Based on the above arguments, we discuss the origin of the ultrafast reduction of the AHE within 100 fs, focusing on the well-defined cluster-octupole phase at 220 K because calculating the band structure in the helical phase is difficult. We consider $T_e$ to be elevated to 700 K by the pump whereas the lattice and spin systems remain unperturbed within 100 fs. This means that photoexcitation does not modify the band structure and only broadens the electron distribution, which allows us to calculate the intrinsic AHE by using Eq. (1). The curves in Fig. 4(d) show the results of the calculated $\sigma_{xy}^{\text{int}}$ normalized at the value of 220 K as a function of $T_e=T_{\text{DFT}}/5$. The different colors correspond to different chemical potentials. The composition of Mn in our sample indicates that the chemical potential is close to zero. The markers in Fig. 4(d) show the experimental results of $\sigma_{xy}$ at 220 K for $I_p$=160, 300, and 500 μJ cm$^{-2}$, corresponding to $T_e$=430, 560, and 700 K, respectively. The result shows that the photoinduced ultrafast decrease of AHE is reasonably accounted for by the calculated intrinsic AHE with the elevated $T_e$. The sensitiveness of the intrinsic AHE to the rise of $T_e$ can be understood from Eq. (1). Nominally all the electron states below $E_F$ can contribute to the intrinsic AHE. However, the Berry curvature are mainly located at around the (anti)crossing points in the band structure and the upper and lower branches of each (anti)crossing point have Berry curvatures with opposite signs as schematically shown in Fig. 4(e). The contribution of the states much below $E_F$ is mostly cancelled by the integration in Eq. (1), and the (anti)crossing points near $E_F$ is dominant. Therefore, the rise in $T_e$ to several hundred K can largely modify the intrinsic AHE and dominates the ultrafast dynamics.

In summary, we presented ultrafast time-resolved study of AHE in Mn$_3$Sn. Isolating the electron temperature from the lattice and spin subsystems, we clearly resolve ultrafast suppression of AHE due to the rise of electron temperature before the photoexcited demagnetization occurs, which is well explained by the intrinsic Berry-curvature mechanism. Importantly, photoexcitation in Mn$_3$Sn largely suppressed the AHE whereas the scattering rate was almost unchanged, which also excludes the extrinsic mechanism. Because our time-resolved study can observe $\sigma_{xx}(\omega)$, $\sigma_{xy}(\omega)$, and the scattering rate with drastically elevating the electron temperature, this method can be utilized to distinguish the microscopic origin of AHE and is in principle applicable to other conventional ferromagnets or topological magnets. This work opens a new avenue for the study of AHE from the viewpoint of nonequilibrium.

**Acknowledgement**

This work was supported by JSPS KAKENHI (Grants Nos. JP19H01817, JP19H00650, JP20J01422, and JP21K13858) and by JST PRESTO (Grant No. JPMJPR20LA), and JST Mirai Program (JPMJMI20A1). Near-infrared transmission and reflection spectroscopy was performed using the facilities of the Materials Design and Characterization Laboratory in the Institute for Solid State Physics, the University of Tokyo. The work at the Institute for Quantum Matter, an Energy Frontier Research Center was funded by DOE, Office of Science, Basic Energy Sciences under Award \# DE-SC0019331. R.M. conceived this project. T.H., H.P., T. Matsuo, and S.N. fabricated the sample and characterized it in the DC measurement. T. Matsuda performed the THz spectroscopy experiments and analysis with helps of N.K., Y.H., N.Y., R.S., and R.M. T.K. conducted the DFT calculation. All the authors discussed the results. T. Matsuda and R.M. wrote the manuscript with substantial feedbacks from all the coauthors.

**Figure captions**

**Fig. 1** (a) Schematics of the atom configuration (up) and the spin texture (down) in Mn$_3$Sn. (b) Temperature dependence of the anomalous Hall resistivity $\rho_{xy}$ in DC transport. (c) A schematic of the optical pump and polarization-resolved THz probe setup. WGP: wire-grid polarizer. (d),(e) THz longitudinal and anomalous Hall conductivity spectra, $\tilde{\sigma}_{xx}(\omega)$ and $\tilde{\sigma}_{xy}(\omega)$, respectively, at 280 K.

**Fig. 2** (a) $x$- and $y$-components of the THz pulse waveforms after transmittance by the sample at 220 K. The vertical line defines $t_{\text{probe}} = t_0$ at the peak. (b) Pump-induced change of $E_y$ at $t_{\text{probe}} = t_0$ as a function of $t_{\text{pump}}$. The data for different pump fluences are shown with offset. The solid curves show results of fitting. (c) Re$\tilde{\sigma}_{xy}(\omega)$ at $t_{\text{pump}}$=0.52 ps for different pump fluences. (d) Re$\tilde{\sigma}_{xx}(\omega)$ at $t_{\text{pump}}$=0.52 ps for the maximum pump fluence of 500 μJ cm$^{-2}$.

**Fig. 3** (a) Temperature dependence of Re$\tilde{\sigma}_{xy}(\omega)$ in THz Faraday rotation. The open circles (ZFC) and closed ones (FC) are the results in zero-field cooling and at ±2 T. (b) Schematics of the spin configurations. (c) Pump-induced change of $E_y$ at $t_{\text{probe}} = t_0$ as a function of $t_{\text{pump}}$ at 150 K. The data for different pump fluences are shown with offset. The solid curves show results of fitting. (d)(e) The change of Re$\tilde{\sigma}_{xy}(\omega)$ at $t_{\text{pump}}$=0.52 and 12 ps, respectively, at 150 K.

**Fig. 4** (a) The electron and lattice temperatures ($T_\text{e}$ and $T_\text{L}$) in the two-temperature model at the original temperatures of 220 and 150 K and $I_\text{p}$=500 μJ cm$^{-2}$. (b) The local spin moment in the DFT calculation with finite temperatures. (c) (left) The band structure of

Mn$_3$Sn at the cluster-octupole phase calculated by DFT. (right) The projected density of states for the spin configuration of the cluster octupole. (d) The circles show Re$\tilde{\sigma}_{xy}(\omega)$ observed in the pump-probe experiment at $t_{\text{pump}}$=0.52 ps at 220 K as a function of the maximum $T_e$. The curves show the calculated $\sigma_{xy}^{\text{int}}$ in Eq. (1) with various chemical potentials as a function of $T_e$. (e) (left) A schematic of the Berry curvature around anticrossing points. (right) Occupation of electrons with low (blue) and high (red) temperatures.

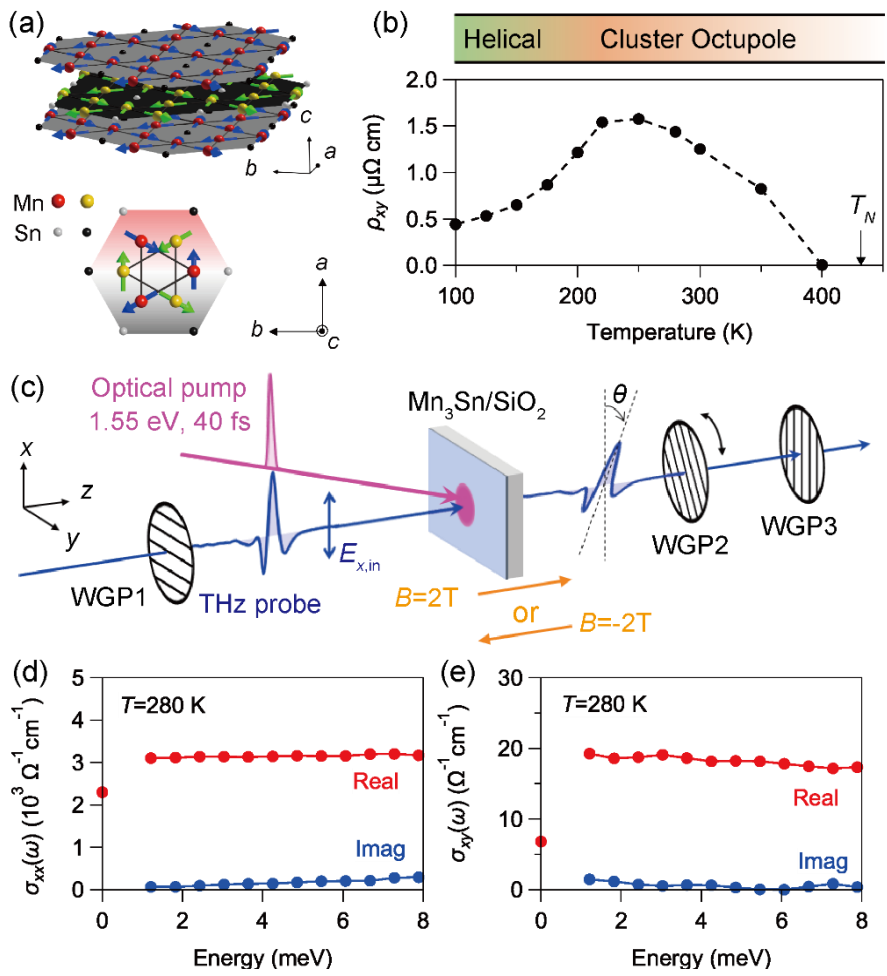

Figure 1

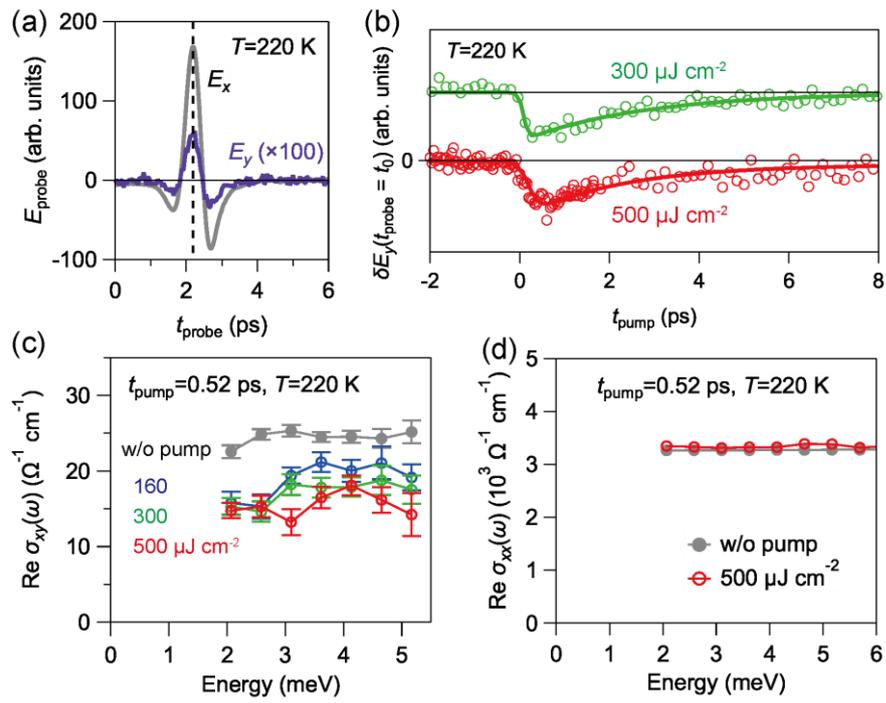

Figure 2

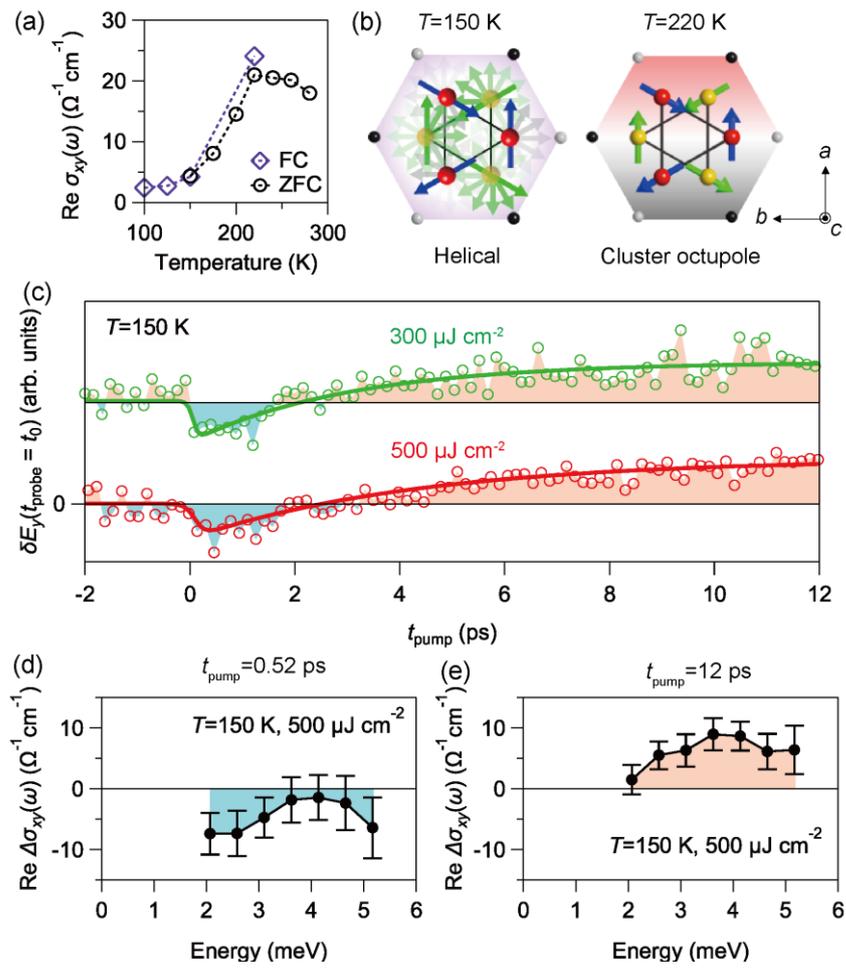

Figure 3

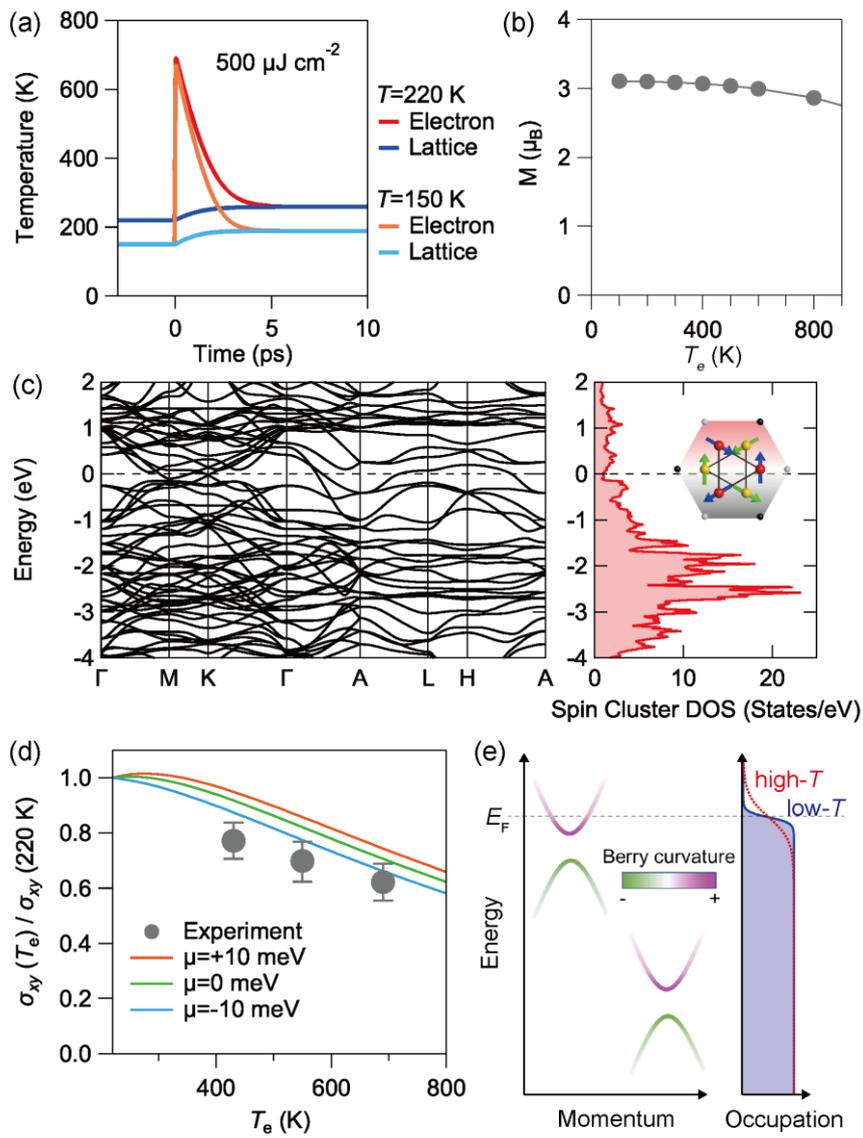

Figure 4